\documentclass[aps, prd, floatfix, nofootinbib, superscriptaddress, twocolumn]{revtex4-1}

\pdfoutput=1

\usepackage{latexsym}
\usepackage{amsmath}
\usepackage{amssymb}
\usepackage{amsfonts}
\usepackage{textcomp}
\usepackage{xcolor}
\usepackage{multirow}
\usepackage{colortbl}
	
\usepackage{float}
\usepackage{color}
\usepackage{supertabular}
\usepackage{placeins}
\usepackage{epsfig}
\usepackage{graphicx}
\usepackage{multirow}

\def\1eq#1{Eq.~(\ref{#1})}

\def\2eqs#1#2{Eqs.~(\ref{#1}) and~(\ref{#2})}
\def\3eqs#1#2#3{Eqs.~(\ref{#1}),~(\ref{#2}) and~(\ref{#3})}

\def\ket#1{\left\vert{#1}\right\rangle}
\def\bra#1{\left\langle{#1}\right\vert}

\begin{document}

\title{A Tailor-made Quantum State Tomography Approach}

\date{\today}

\author{Daniele Binosi}
\email{binosi@ectstar.eu}
\affiliation{European Centre for Theoretical Studies in Nuclear Physics
and Related Areas (ECT*, Fondazione Bruno Kessler); Villa Tambosi, Strada delle Tabarelle 286, I-38123 Villazzano (TN), Italy}

\author{Giovanni Garberoglio}
\email{garberoglio@ectstar.eu}
\affiliation{European Centre for Theoretical Studies in Nuclear Physics
and Related Areas (ECT*, Fondazione Bruno Kessler); Villa Tambosi, Strada delle Tabarelle 286, I-38123 Villazzano (TN), Italy}

\author{Diego Maragnano}
\email{diego.maragnano01@universitadipavia.it}
\affiliation{Dipartimento di Fisica, Universit\`a  di Pavia, via Bassi 6, 27100 Pavia, Italy}

\author{Maurizio Dapor}
\email{dapor@ectstar.eu}
\affiliation{European Centre for Theoretical Studies in Nuclear Physics
and Related Areas (ECT*, Fondazione Bruno Kessler); Villa Tambosi, Strada delle Tabarelle 286, I-38123 Villazzano (TN), Italy}
\author{Marco Liscidini}
\email{marco.liscidini@unipv.it}
\affiliation{Dipartimento di Fisica, Universit\`a  di Pavia, via Bassi 6, 27100 Pavia, Italy}

\begin{abstract}
Quantum state tomography (QST) aims at reconstructing the state of a quantum system.  However in conventional QST the number of measurements scales exponentially with the number of qubits. Here we propose a QST protocol, in which the introduction of a threshold allows one to drastically reduce the number of measurements required for the reconstruction of the state density matrix without compromising the result accuracy. In addition, one can also use the same approach to reconstruct an approximated density matrix tailoring the number of measurements on the available resources.
We experimentally demonstrate this protocol by performing the tomography of states up to 7 qubits. We show that our approach can lead to results in agreement with those obtained by QST even when the number of measurements is reduced by more than two orders of magnitude.
\end{abstract}

\maketitle


\section{Introduction}
In the last few years, the race for quantum technologies has favoured the realization of large and complex quantum states in various platforms, including superconducting {\cite{cao2023generation}}, atomic {\cite{haffner2005scalable, 14ions}}, and photonic systems {\cite{10_photon_entanglement, zhang2019generation,10_qubit_cat_state,photon_qudits_boyd,one_way_qp_photons,xanadu_borealis}}. Such advances are an important resource for the implementation of useful quantum computers and an extraordinary opportunity for fundamental studies in quantum mechanics. 

Ideally, one would like to know the complete state of a quantum system, for this is sufficient to compute the value of any other observable. However, as the dimension of the quantum system increases, this task is not trivial at all. For example, consider quantum state tomography (QST) \cite{james_2001}, which is arguably the most famous approach for quantum state reconstruction. The idea is to obtain the density matrix of the system from the measurements of a set of observables on a large enough ensemble of identical copies of the state. This method is general and does not make any \emph{a priori} assumption on the state. However, a system composed of $n$ qudits requires determining the expectation value of at least $d^{2n}$ observables, that is the number of independent entries of the density matrix. Such an exponential growth of the number of measurements with the number of qudits can make the experimental implementation of QST unfeasible even for just three or four qudits, depending on the quantum system under investigation. 

Over the years, strategies have been developed to decrease the number of required measurements, usually by leveraging certain assumptions about the state or the anticipated outcome. For example, the efficiency of compressed sensing QST depends on the rank of the density matrix, with the state characterization done via a certain number of random Pauli measurements {\cite{gross_2010}}. Further improvements of this approach have reduced the number of necessary measurement sets significantly \cite{cs2019a, cs2019b}. one necessitates the state to be uniquely determined by local reduced density matrices or the assumption that the global state is pure \cite{qst_via_rdms}.  In Bayesian QST, one defines conditional probabilities starting from measurement results and a suitable \emph{a priori} distribution over the space of possible states {\cite{derka_bayesian, lukens_bayesian_natcomm}}. In self-guided QST one addresses tomography through optimization rather than estimation, iteratively learning the quantum state by treating tomography as a projection measurement optimization problem \cite{Ferrie_2014,Chapman_2016,Rambach_2021}.

\begin{figure*}[!t]
\centering
\includegraphics[width=15.cm]{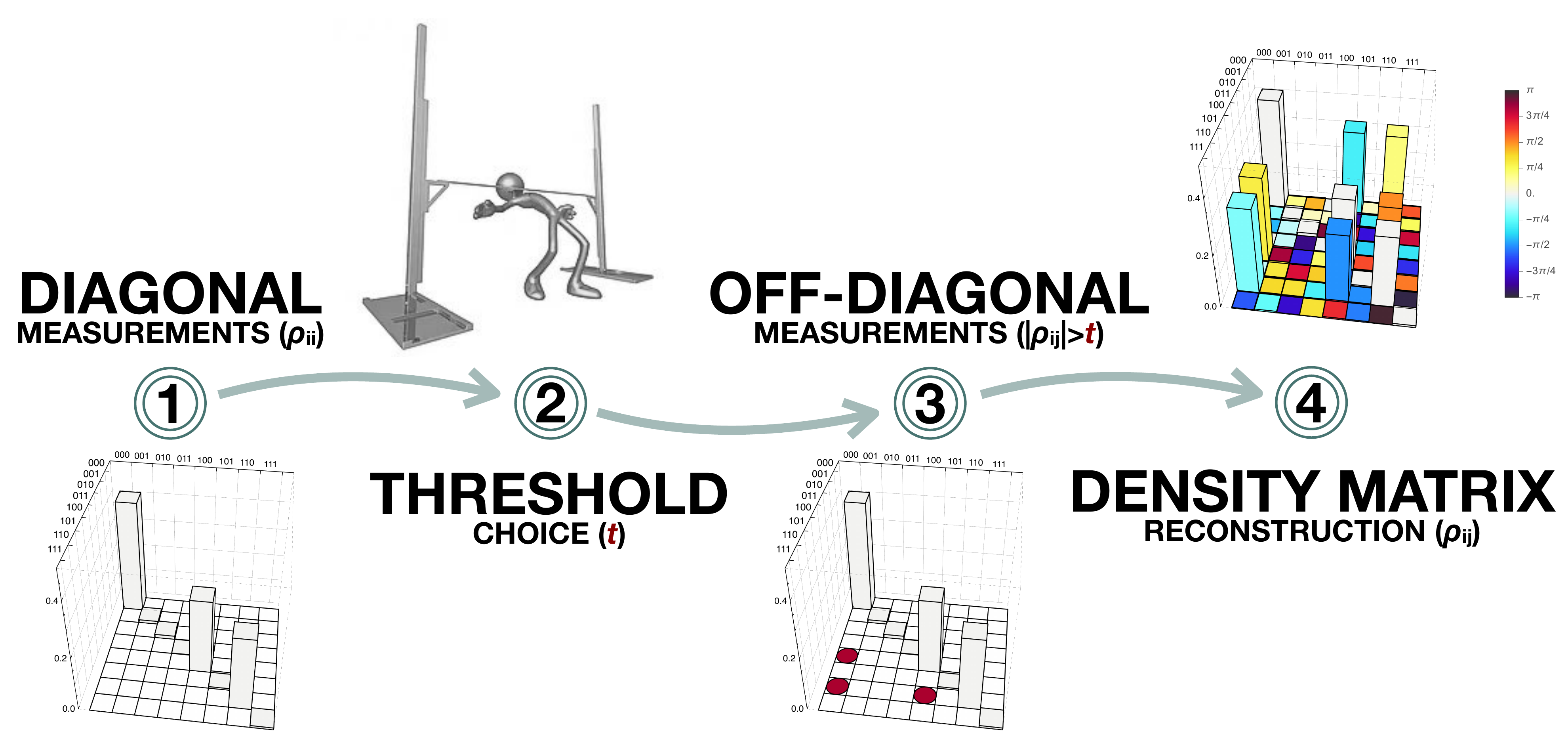}
\caption{\label{fig1:protocol}Pictorial representation of the tQST process. (1)  Measurement of the diagonal terms $\rho_{ii}$; (2) Choice of the threshold $t$. The choice of $t$ resembles that of positioning the bar in a limbo competition. If the bar is very low ($t\ll 1$), a successful performance will certainly be spectacular. However, if the bar is too low, the dance may be prohibitive even for a most skilled athlete. This is the situation of conventional QST when the number of measurements to be performed is unreasonable. On the contrary, for a sufficiently high bar, even a beginner may be able to dance, but the result may not be particularly noteworthy. (3) Construction and measurement of the observable associated with the off-diagonal terms; (4) Reconstruction of the density matrix; we plot the absolute values of the matrix elements colored by their phase as indicated by the color bar.}
\end{figure*}

Other strategies avoid the reconstruction of the density matrix by focusing only on \emph{some} relevant information about the state. For instance, in shadow tomography, one can estimate the value of a large number of observables with a few copies of the unknown state {\cite{aaronson_shadow, huang2020predicting, experimental_shadow}}. Finally, quantum witnesses are useful to verify some crucial properties of a quantum state, such as its degree of entanglement and the kind of quantum correlation {\cite{min_tomo_ent_witnesses}}. While these last two approaches are extremely useful and applicable to states of large dimensions, it is somewhat disappointing that, after all the efforts to implement a state, one cannot look at it as a whole. 

In this work, we describe and analyze \emph{threshold} Quantum State Tomography (tQST)~\cite{icton2023}, a protocol that allows one to reconstruct the density matrix of a quantum state, even of large dimension, by allowing an efficient trade-off between the number of measurements and the accuracy of the state reconstruction through the presence of a \emph{threshold} parameter. We show that tQST can drastically reduce the resources required for state reconstruction, but it can also be used to obtain an approximated density matrix by further reducing the number of measurements and the experimental efforts. The protocol can be applied to any quantum system and does not make any assumption about the state to be characterized. First, we outline the tQST procedure and discuss its implementations in the case of $n$-qubits. Second, we experimentally verify the protocol by performing tQST on systems up to 7 qubits with the IBMQ platform and compare the results with conventional QST. Finally, we draw our conclusions.

\section{\label{sec:Results}Results}
\subsection{The threshold quantum state tomography protocol}
In conventional QST, the number of measurements required to reconstruct the density matrix is uniquely determined by its dimension \cite{james_2001}. Specifically, a system of $n$ qubits requires at least $4^{n}$ measurements, that is the number of independent parameters of the state density matrix. These measurements can be performed in an arbitrary sequence, and the results are finally combined to obtain the state density matrix through maximum likelihood estimation or other approaches \cite{hradil, numerical_strategies, Lukens_2020}.

\begin{figure*}[!t]
\centering
\includegraphics[height=4.5cm]{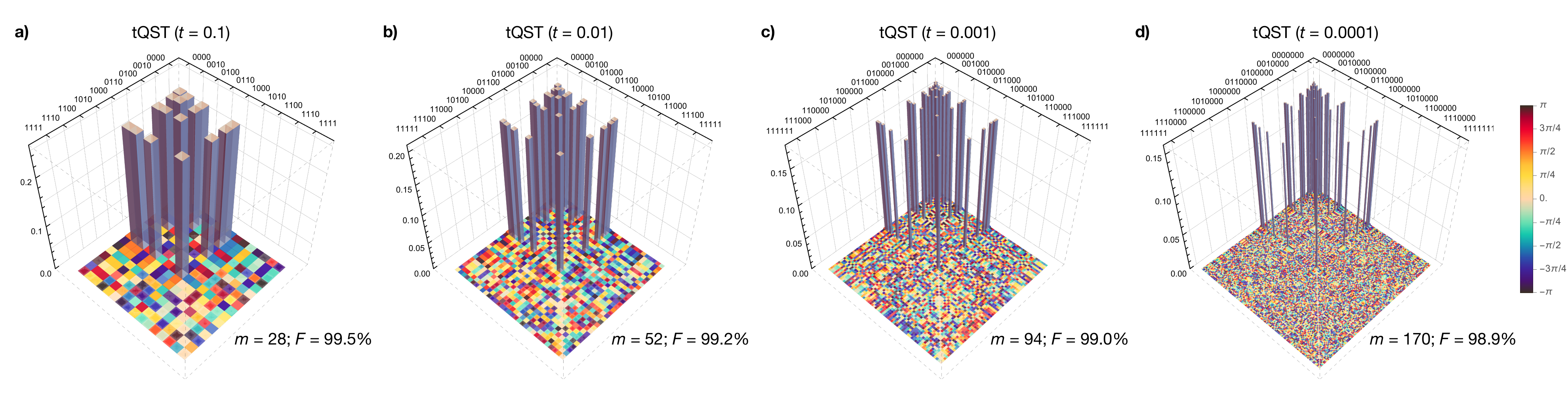}
\caption{\label{fig2:}Reconstruction of the density matrix of a W state via tQST. We consider different numbers of qubits and threshold values: a) 4 qubits, $t=0.1$ b) 5 qubits, $t=0.01$, c) 6 qubits, $t=0.001$, and d) 7 qubits, $t=0.0001$. For each case, we indicate the number of measurements used and the fidelity with the corresponding target state.}
\end{figure*}

Here, we begin by noting that a density matrix, denoted as $\rho$, is required to have a unit trace, be Hermitian, and positive semi-definite. Specifically, this latter property entails that any element $\rho_{ij}$ of a physical density matrix $\rho$ must fulfill the condition $|\rho_{ij}|\le\sqrt{\rho_{ii}\rho_{jj}}$. Thus, measuring the diagonal elements of the density matrix ({\it i.e.}, projecting on the states of the computational basis) provides some information on the off-diagonal terms. Indeed, if $\rho_{ii}$ is found to be zero, then one can immediately set to zero all the elements of the $i$-th row and column of $\rho$. Similarly, if $\rho_{ii}$ and $\rho_{jj}$ are different from zero but small compared to the other diagonal elements, one knows that the modulus of $\rho_{ij}$ will be small too. 

These considerations are at the basis of tQST, whose protocol is illustrated in Fig.~\ref{fig1:protocol}. First, one measures the diagonal elements $\{\rho_{ii}\}$ of the density matrix, which are directly accessible by projecting on the chosen computational basis. Second, one chooses a threshold $t$ and, by exploiting the information on $\{\rho_{ii}\}$, identifies those $\rho_{ij}$ for which $\sqrt{\rho_{ii}\rho_{jj}}\ge t$. Third, one constructs a proper set of observables associated with these $\rho_{ij}$ (see Methods) and performs only those measurements. Finally, one uses these results to reconstruct the density matrix, for example, using a maximum-likelihood estimation.

In tQST, the information achieved by measuring the diagonal terms of $\rho$ is immediately used to decide the subsequent measurements to be performed, by choosing to neglect the terms of $\rho$ that in modulus are smaller than a certain threshold. Unlike conventional QST, the resources necessary to reconstruct the state are not uniquely determined by the dimension of the quantum system but can be controlled with the threshold $t$. For example, if one sets $t=0$, all the elements of $\rho$ are considered, and the protocol reduces to conventional QST. On the contrary, for $t>0$, the protocol may require fewer measurements than those needed with conventional QST and, in any case, no more than them. Importantly, one does not make any \emph{a priori} assumption on the state or the result of the characterization.  It should be noted that unlike adaptive approaches \cite{cs2019a,cs2019b}, in which the necessary measurement sets cannot be known \emph{a priori}, for each one is chosen based on the outcomes of the previous measurement set, in tQST the projectors can all be determined once the system is measured in the chosen computational basis. In this respect, we hasten to emphasize that,  once a threshold value is chosen, the protocol does not simply reduce to measuring a subspace of the Hilbert space of the whole system.

The threshold $t$ determines the amount of information one is willing to trade in exchange for fewer measurements. This has several consequences. First, one may be able to reconstruct the entire density matrix of a large state by reducing the amount of measurements to a level compatible with the available experimental resources. Second, fewer measurements can give a significant advantage in terms of storing and handling experimental data. Third, one may be able to avoid useless measurements and increase the integration time for the remaining measurements, leading to an improvement of the signal-to-noise ratio. The amount of information obtainable in the characterization of a quantum state is always limited, for example by noise or the finite precision of the experimental setup.
Such experimental constraints \emph{de facto} bound the accuracy with which $\rho$ can be determined, even for traditional QST. Thus, although one may naively expect that reducing the number of measurements will decrease the quality of the results, as we shall see in the following, a wise choice of $t$ can still guarantee practically reaching the best achievable result while requiring fewer resources. 

In Fig.~2 we show the simulated reconstruction of the density matrix via tQST with maximum likelihood estimation of several W states for different numbers of qubits ranging from 4 to 7~\cite{w_states}. Even for the case of 7 qubits, in which traditional QST would have required 16,384 measurements, one can reconstruct $\rho$ with a fidelity of about 99\% with the target state by implementing only 170 measurements. We stress that here the dramatic reduction of measurements is not simply given by the choice of a particular state, but rather by the employed computational basis, which affects the final state representation. As in compressed sensing tomography or similar approaches, sparse matrices are usually easier to reconstruct, because most of the information is contained in fewer elements of the density matrix. Yet, for sparse matrices the advantage of tQST can be significantly larger than with other approaches. Take for example the 7-qubit state reconstructed in Ref. \cite{riofrio2017experimental} via compressed sensing QST, in which the characterization required 16,256 projective measurements. While this number is significantly smaller than $6^n$, which is the number of observables of the typical tomographically overcomplete set, this is still 99\% of the $4^n$ observables that are strictly necessary \cite{james_2001}. On the contrary, the very same state could be reconstructed via tQST by performing only 184 measurements, which is about 1\% of those required by compressed sensing (see Supplementary Material). We stress that tQST can also be applied to matrices that are not sparse, where the threshold $t$ will determine both the number of measurements and the reliability of the reconstruction. 

As shown in Fig.~2, a notable feature of tQST is that a significant reduction in the number of measurements does not necessarily lead to large errors in the reconstruction of the density matrix. However, this feature is strongly dependent on the state representation.  More in general, one may be interested in using tQST because the available resources are simply not enough to implement  the traditional QST. In this case, it is important to have an estimate of the largest error that is associated with the threshold choice. Specifically, given the diagonal elements $\{\rho_{ii}\}$ and a corresponding threshold $t$, one can estimate a lower bound for the fidelity achievable through a tQST reconstruction (see Sec.~\ref{sec:fbound}):
\begin{eqnarray}\label{fbound}
    F_\mathrm{bound}(\{\rho_{ii}\},t)=\left(
    1-\sqrt{r(\bar{\rho})\sum_{\{ij\} \in \mathcal{C}_t} \rho_{ii} \rho_{jj}}
    \right)^2,
\end{eqnarray}
where $\rho_t$ is an estimator matrix defined according to \mbox{$\rho_t=\{0 \quad \forall\ \{ij\}\in{\cal C}_t; \sqrt{\rho_{ii}\rho_{jj}} \ {\mathrm{otherwise}}\}$}] with $\mathcal{C}_t$ the ensemble of the elements $\rho_{ij}$ that are below threshold ({\it i.e.}, $\mathcal{C}_t = \left\{ \{ij\} \mid \sqrt{\rho_{ii} \rho_{jj}} < t \right\}$), and $r \left( \bar{\rho} \right)$ is the rank of the ideal density matrix.
We stress that in some cases this can be a quite conservative lower bound but still useful to verify whether a resulting low fidelity depends on the choice of an excessively high threshold,  or rather a real dissimilarity between $\rho$ and the target state.

We developed a Python package which implements tQST for an arbitrary number of qubits that is freely available to users on Github \cite{Binosi_tresholdqst_A_Python_2024}.

\begin{figure*}[!t]
\centering
\includegraphics[height=19cm]{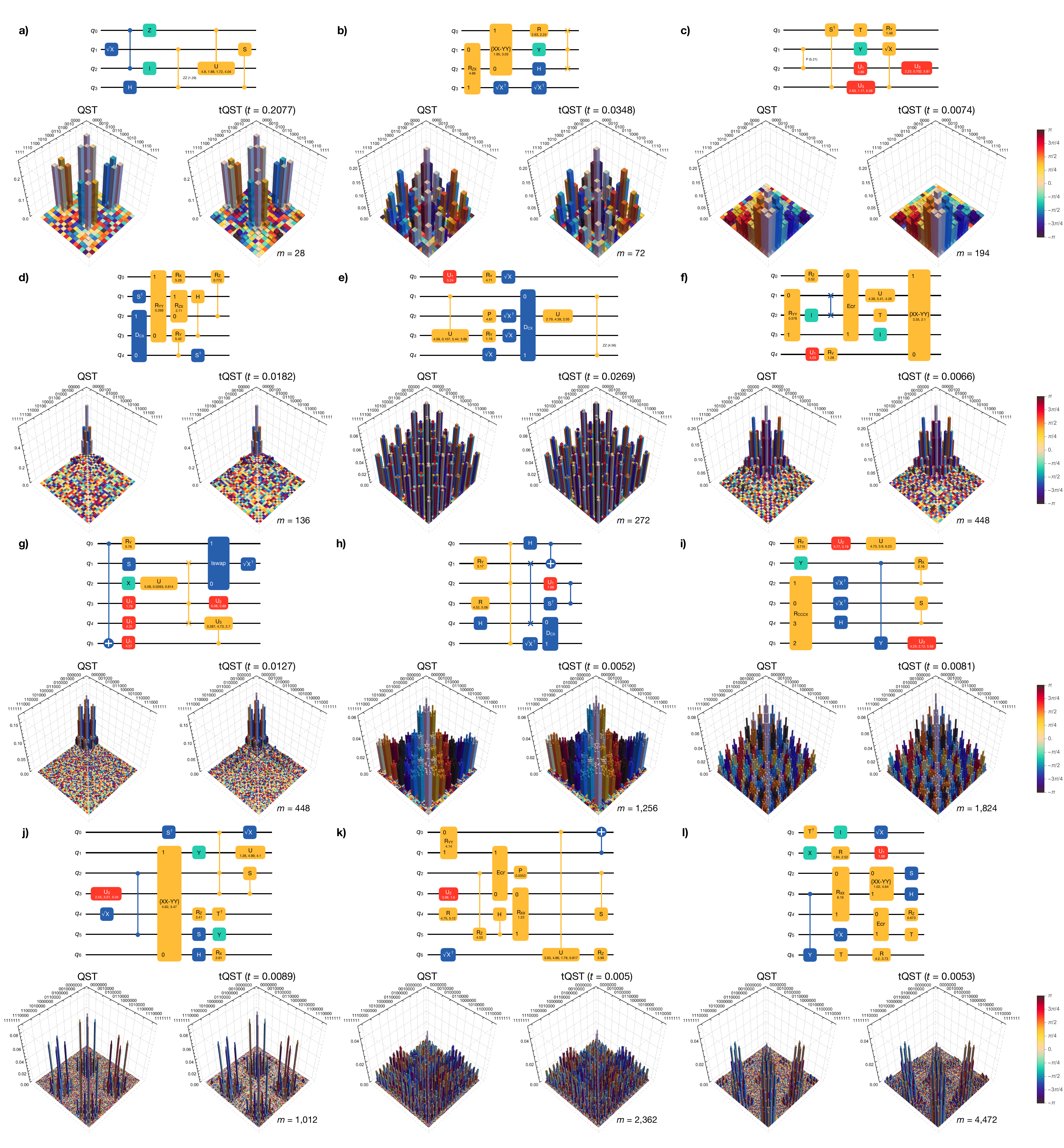}
\caption{\label{IBMQ}tQST on random circuits implemented in a IBMQ processor. a), b), c): 4-qubit states generated with the shown circuits with a diagonal filling of 25\% , 50\%, and 75\%, respectiv4ely.
d) - l): same as before, with 5, 6, and 7 qubits.
}
\end{figure*}

\begin{figure*}[!t]
\centering
\mbox{}\hspace{-1.75cm}\includegraphics[width=18.5cm]{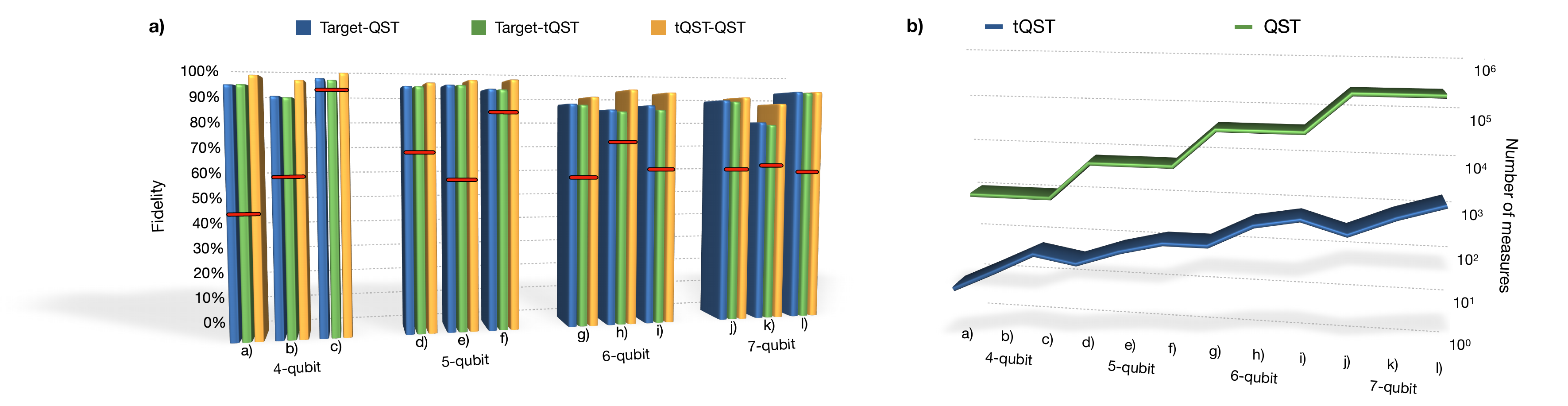}
\caption{\label{Fidelity}Quantitative comparison between the reconstructed QST and tQST density matrices of the random circuits illustrated in Fig.~\ref{IBMQ}. a) Comparison of QST and tQST density matrix fidelities. The red line represents the tQST fidelity bound corresponding to the specified circuit threshold. b) Scaling of the required measurements with the number of qubits for both QST and tQST reconstructions.}
\end{figure*}


\subsection{Implementation on an IBMQ processor}




We experimentally demonstrate tQST by using the IBMQ processor \texttt{lagos} \cite{Qiskit, kanazawa2023qiskit}, which allows one to prepare states with up to 7 superconducting qubits. 
In our implementation, each time we program the system to generate a target quantum state by constructing the corresponding quantum circuit. We first reconstruct the state density matrix by performing conventional QST as implemented by IBMQ, which uses linear inversion on the outcomes of an overcomplete set of $6^n$ observables. This process generally yields a non-positive reconstructed state, which is subsequently rescaled to be positive semi-definite using the method outlined in~\cite{Smolin:2011tob}. We then reconstruct the same state by using tQST, where the threshold $t$ is chosen by considering the typical signal-to-noise ratio (SNR) of the system to avoid unnecessary measurements (Methods, Sec. \ref{subsec:threshold}). In this case, we use a maximum likelihood estimation to obtain the (positive semi-definite) density matrix from the measured expectation values (Methods, Sec. \ref{subsec:ml}). To compare the results we compute the fidelities between the target state and the reconstructed ones, and also that between the two reconstructed states. This choice is motivated by the fact that while not exhaustive, fidelity reamins a widely recognized metric. 

In our analysis, we generated several random states and sorted them according to the diagonal filling, {\emph i.e.}, the percentage of expected non-vanishing diagonal elements. In Fig.~\ref{IBMQ} we show representative results for filling percentages of 25\% (a,d,g,j), 50\% (b,e,h,k), and 75\% (c,f,i,l) for a number of qubits ranging from 4 (a,b,c) to 7 (j,k,l). Each panel of the figure shows the circuit to generate the state, the density matrix reconstructed with conventional QST, and the one obtained with tQST. In this last case, we also report the value of the threshold $t$ and the corresponding number of measurements used in the reconstruction. Since we choose $t$ according to the typical SNR of the system, such a number is also related to the sparsity of the state representation. For the cases of 4, 5, and 6 qubits, we observe up to a 100-fold reduction in the number of required measurements compared to IBMQ-QST. For the case of 7 qubits, such reduction is $\sim 300$ for the case of Fig.~3j. 

As evident from the state representation, despite the significant difference in the number of measurements, the state reconstructed with IBMQ-QST and tQST are very similar. A more quantitative analysis is obtained by calculating the fidelity between the two reconstructed states, which in all cases is about 90\% or above, as shown in Fig.~\ref{Fidelity}. The same figure also reports the fidelity of the tQST-reconstructed state with the target one. Remarkably, this is always comparable (within errors) to the fidelity between the IBMQ-QST reconstructed state and the target one (see Supplementary Materials, Tab. S1). Thus no advantage is obtained with IBMQ-QST by performing more measurements than those set by the threshold in tQST. This suggests that when the threshold is determined by the SNR of the system,  the tQST protocol can extract all the amount of information accessible with conventional QST, yet with a smaller number of measurements.

The IBMQ processor \texttt{lagos} limits our analysis to 7 qubits. Yet, it is interesting to further investigate the performance of tQST by extrapolating the analysis to a larger number of qubits. In this analysis we considered simulated data with a SNR analogous to that of the IBMQ system and followed the same strategy for the threshold choice. We considered the case of W states and increased the number of qubits up to~14, limited now only by our hardware. In Table \ref{tab:my_label} we report the number of qubits, the threshold $t$, the number of measurements, and the fidelity with respect to the target state. In all the cases we obtained fidelities exceeding 90\%. We stress that in the cases of 14 qubits, the 16,556  measurements required by tQST make the reconstruction in principle experimentally accessible today. On the contrary, conventional QST would necessitate at least $4^{14}=268,435,456$ measurements that, at the moment, appear as a prohibitive number.

\begin{table}[!t]
    \centering
    \begin{tabular}{r||r|r|r}
          Qubits\hspace{0.2cm}  & 
         \hspace{0.2cm}Threshold ($t$)\hspace{0.2cm} &
         \hspace{0.2cm}Measures\hspace{0.2cm} & \hspace{0.2cm}Fidelity\hspace{0.2cm} \\
         \hline\hline
         8 & 0.053 &   312 & 91.5\%  \\
         9 & 0.047 &   584 & 91.9\%  \\
        10 & 0.042 &  1,114 & 91.2\%  \\
        11 & 0.038 &  2,158 & 91.4\%  \\
        12 & 0.035 &  4,228 & 91.4\%  \\
        13 & 0.032 &  8,348 & 91.3\%  \\
        14 & 0.030 & 16,556 & 91.3\%  \\
        \hline
    \end{tabular}
    \caption{tQST reconstruction of a W state from synthetic data where an {\em ad-hoc} noise similar to the one observed in Ref.~\cite{haffner2005scalable} has been introduced. 
    The number of the needed off-diagonal measures for the state reconstruction scales as $n^2-n$ with the number of qubits, and therefore it is subdominant with respect to the number of diagonal measurements ($2^n$) necessary for the determination of the threshold.}
    \label{tab:my_label}
\end{table}

\section{Discussion}
We have proposed and implemented a protocol for QST that allows one to reconstruct the density matrix of any quantum state with a number of measurements that can be considerably smaller than that required by conventional QST. Such an approach does not make any \emph{a priori} assumptions on the state and employs a threshold $t$ to control the amount of information used in the state reconstruction. The presence of such a threshold allows one to trade the minimum fidelity with which the state can be reconstructed with the amount of resources to perform the characterization, thus reducing the number of measurements significantly. In addition, the threshold can be set to take into account the experimental limitations that may lead to unnecessary measurements. Our protocol was implemented on the IBMQ system \verb|lagos| to characterize random states up to 7 qubits. In all the considered cases the fidelity achieved with tQST was compatible within the experimental uncertainty with that obtained by using conventional QST but with a smaller amount of measurements (in some cases $\sim300$ times smaller). This suggests that our protocol is able to efficiently access all the information that can be extracted from the system. Finally, by using synthetic data, we pushed the approach to our computational limit and performed the characterization of W states up to 14 qubits, reaching a fidelity larger than 90\% with only $\sim16,000$ expectation values, four orders of magnitudes less than what would be required by conventional QST. 

Our protocol is a flexible and practical approach for the full characterization of large quantum systems, including those based on atoms or photons. For this reason, we believe that it will be particularly useful to the whole quantum community. 

\section{Materials And Methods}

\subsection{$n$-qubit projectors}
\label{sec:projectors}

The real (respectively, imaginary) part of an off-diagonal element of a density matrix $\rho$, can
be obtained from:
\begin{align}
	\rho^\mathrm{re,im}_{ij}=\mathrm{tr}\left({\cal O}^\mathrm{re,im}_{ij}\rho\right),
\end{align}
where ${\cal O}^\mathrm{re}_{ij}$ (respectively, ${\cal O}^\mathrm{im}_{ij}$) is an operator with:
$1/2$ (respectively, $i/2$) at the entry $ij$; $1/2$ (respectively, $-i/2$) at the entry $ji$; and
zero otherwise. 
How to experimentally conduct a series of measurements corresponding to these operators remains unknown. Instead, one can determine a set of $4^n$ projectors, denoted as $P=\ket{\psi}\bra{\psi}$, where $\ket{\psi}$ is separable: $\ket{\psi}=\otimes_{k=1}^n\ket{\phi_k}$ with $\ket{\phi_k}$ an eigenstate of the Pauli matrices.
The projectors associated with the real or imaginary part of the matrix element $ij$ of the density matrix are then given by
\begin{align}
P^\mathrm{re,im}_{ij} = \underset{P}{\mathrm{argmin}} \, \|{\cal O}^\mathrm{re,im}_{ij} - P\|_2,
\label{eq:min}
\end{align}
where $\|{\circ}\|_2$ represents the Frobenius norm.

A given set of projectors is said to be {\em
  tomographically complete} if there exists a unique set of scalars $a^\mathrm{re,im}_{K}$ (with
$K=\{ij\}$ a multi-index) such that the density matrix can be reconstructed, that is
\begin{align}
	\rho=\sum_{K}a^\mathrm{re,im}_{K}P^\mathrm{re,im}_{K}.
\end{align}     
Equivalently, suppressing the ``re'' and ``im'' superindices, one gets
\begin{align}
	\mathrm{tr}\left(P_L^\dagger \rho\right)=\sum_Ka_K\mathrm{tr}\left(P_L^\dagger P_K\right)=\sum_KM_{LK}a_K.
\end{align}
which means that the matrix $M_{LK} = \mathrm{tr}\left(P_L^\dagger P_K\right) = |\langle \psi_L | \psi_K           
\rangle|^2$ has to be invertible and the scalars $a_K$ are uniquely
determined as~$a_K = \left(M^{-1}\right)_{KL}\mathrm{tr}\left(P_L^\dagger \rho\right)$~\cite{james_2001}. Notice that, owing to the presence of equivalent minima in Eq.~(\ref{eq:min}),  it is not guaranteed that a complete set exists; however, if a complete set exists, then the matrix $M$ is positive definite.

We found that in the case of an $n$-qubit system, a tomographically complete set of projectors associated to the elements of a $n$-qubit density matrix can
be obtained from the eigenvectors of the Pauli matrices. In the conventional polarization notation, these eigenvectors are: 
\begin{align}
	&\ket{H};\! & \ket{D}=\frac1{\sqrt2}\left(\ket{H}+\ket{V}\right);&\!\! &\ket{R}=\frac1{\sqrt2}\left(\ket{H}+i\ket{V}\right),& \nonumber \\
	&\ket{V}; &\ket{A} =\frac1{\sqrt2}\left(\ket{H}-\ket{V}\right);& &\ket{L} =\frac1{\sqrt2}\left(\ket{H}-i\ket{V}\right).&
	\label{eq:pol}
\end{align} 
In the 1-qubit case one possibility is: $P^\mathrm{re}_{11}=\ket{H}$; $P^\mathrm{re}_{22}=\ket{V}$; $
P^\mathrm{re}_{12}=\ket{D}$; and $P^\mathrm{im}_{12}=\ket{R}$; or, in a more suggestive form
\begin{align}
	\pi_1 \equiv
	\begin{bmatrix}
    \ket{H} & \ket{D}+i\ket{R} \\
    0 & \ket{V}
\end{bmatrix}.
\label{eq:p1}
\end{align}
This is a $2 \times 2$ table structured in such a way that 
the projector associated with measuring the real or imaginary component of the element $\rho_{ij}$ of the 1-qubit density matrix is given by the real or imaginary part of the entry $(ij)$ in (\ref{eq:p1}).
For example, information on the imaginary part of $\rho_{12}$ is encoded in the matrix element $\bra{R} \rho \ket{R}$.
Entries with a value of ``0'' (which indeed acts as the zero element for the recursive operations below) do not need to be explicitly determined, as the density matrix is Hermitian. Consequently, we will assume, without loss of generality, that $j \geq i$ whenever we aim to determine $P_{ij}$.

A set of tomographically complete separable projectors for $n>1$ can be then constructed through the following recursion relation: 
\begin{align}
	\pi_{n}=
	\begin{bmatrix}
    \ket{H}\pi_{n-1} & \ket{D}\pi_{n-1} + i \ket{R}\overline{\pi}_{n-1} \\
    0 & \ket{V}\pi_{n-1}
\end{bmatrix}.
\label{eq:rec}
\end{align}
where we have defined 
\begin{align}
	\overline{\pi}_1 \equiv
	\begin{bmatrix}
    \ket{H} & 0 \\
    \ket{D}-i\ket{R} & \ket{V}
    \end{bmatrix},	
\end{align}
and analogously for all $n$.
So in the $2$-qubit case, Eq.~(\ref{eq:rec}) in conjunction with the results
\begin{subequations}
\begin{align}
	\ket{X}\pi_1 =
	\begin{bmatrix}
    \ket{XH} & \ket{XD}+i\ket{XR} \\
    0 & \ket{XV}
\end{bmatrix},\\
	\ket{X}\overline{\pi}_1 =
\begin{bmatrix}
    \ket{XH} & 0 \\
    \ket{XD}-i\ket{XR} & \ket{XV}
    \end{bmatrix},	
\end{align}    
\end{subequations}
(with $X=H,V,D,R$), yields
\begin{widetext}
\begin{align}
	\pi_2 =
\begin{bmatrix}
    \ket{HH} & \ket{HD}+i\ket{HR} & \ket{DH}+i\ket{RH} & \ket{DD}+i\ket{DR}\\
    0 & \ket{HV} & \ket{RR}+i\ket{RD} & \ket{DV}+i\ket{RV} \\
    0 & 0 & \ket{VH} & \ket{VD}+i\ket{VR}\\
    0 & 0 &  0 & \ket{VV}
\end{bmatrix}.
\end{align}
\end{widetext}
Information about of, {\it e.g.}, the real (imaginary) part of $\rho_{23}$ is then encoded in the matrix element $\bra{RR}\rho\ket{RR}$ ($\bra{RD}\rho\ket{RD}$). 

The outlined procedure yields a total of $4^n$ projectors, and we have numerically verified that up to $n=14$ the corresponding matrix $M$ is invertible.
In Eq.~(\ref{eq:rec}), the notation $\ket{H} \pi$ signifies a table where the entries result from the product of the ket $\ket{H}$ and the kets contained in $\pi$. Therefore, $\pi_n$ is a table that has twice the number of rows and columns compared to $\pi_{n-1}$.

The recursive structure outlined in Eq.~(\ref{eq:rec}) can be leveraged to reduce the necessity of generating the entire set of projectors upfront. Instead, we can generate projectors on-demand, specifically for elements of the density matrix that need to be measured to achieve faithful reconstruction given a certain threshold.

\begin{figure*}[!t]
\centering
\includegraphics[width=11.cm]{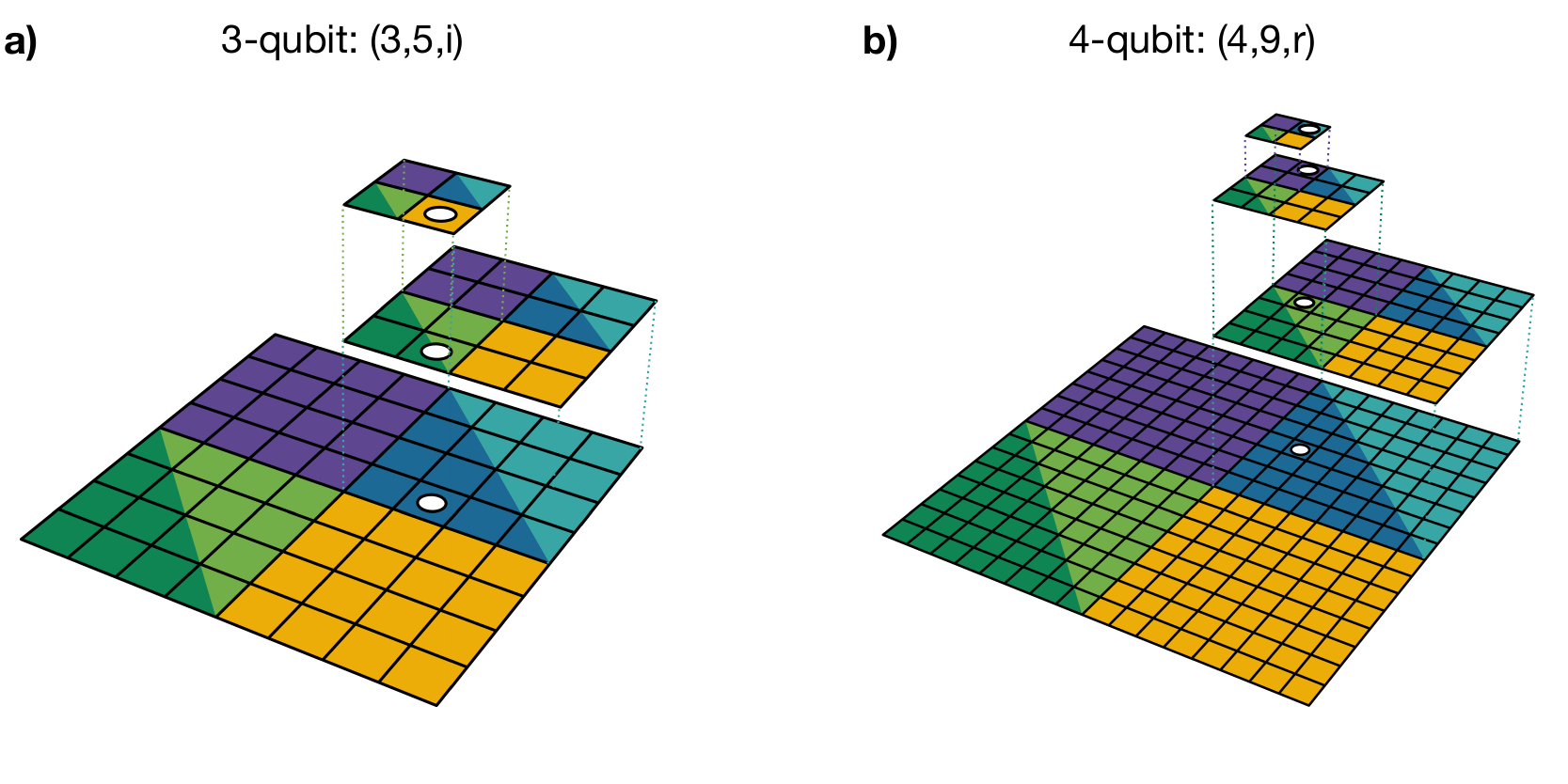}
\caption{\label{fig5:tQST-fig4-high-res}Example determination of the projectors associated to two density matrix elements of 3- and 4-qubit systems using the recursive procedure outlined in the main text. At each state, the density matrix is divided into 4 different quadrants: quadrant ``1'' is colored in violet, quadrant ``2'' in blue, quadrant ``3'' in green, and quadrant ``4'' in golden yellow.
Quadrants ``2'' and ``3'' are further subdivided into an upper (lighter) and a lower (darker) part. 
Each plane represents a successive iteration of the algorithm outlined in the main text, which proceeds from bottom to top.
a) The 3-qubit matrix element $\rho_{3,5}^\mathrm{im}$ is located in quadrants ``$2l3l4$'' and corresponds to the $\ket{RDV}$ projector according to Tab.~\ref{tab:rec} b) 4-qubit matrix element $\rho_{4,9}^\mathrm{re}$ is located in quadrants ``$2l3u12u$'' therefore corresponding to the projector $\ket{RRHD}$ according to Table~\ref{tab:rec}.} 
\end{figure*}

To this end, we divide $\pi_n$ into four quadrants, with ``1'' referring to the upper-left quadrant and ``4'' indicating the lower-right quadrant. Each of the quadrants 2 and 3 is further divided into an upper (``$u$'') and a lower (``$l$'') triangular part. The real part of the elements along the diagonal is assigned to the $u$ portion, while the imaginary part is assigned to the $l$ portion. This subdivision is pictorially represented in Fig.~\ref{fig5:tQST-fig4-high-res}.

To determine the projector corresponding to the density-matrix element $\rho_{i,j}$, we initially locate it within $\pi_n$ which immediately determines the projector associated with the first qubit, that is $\ket{H}$, $\ket{D}$, $\ket{R}$, or $\ket{V}$ according to its position: 1, $2u$, $2l$, or 4, respectively. We then continue splitting the quadrant where the element is found until we reach a resulting quadrant size of $2\times2$. At each splitting step, if the element falls into quadrants 1 or 4, the projector associated to the next qubit is $\ket{H}$ (for 1) or $\ket{V}$ (for 4). Conversely, if it falls into quadrants 2 or 3, the projector choice depends on 
its position in the previous splitting step. 
For an element in an upper quadrant, we select $\ket{D}$; and, for an element in a lower quadrant, we select $\ket{R}$ unless the previous quadrant was either $2l$ or $3l$, in which case the choice is reversed.
Table~\ref{tab:rec} summarizes these steps.

As an illustration, Fig.~\ref{fig5:tQST-fig4-high-res} shows the determination of two projectors for a 3- and a 4-qubit system.
More specifically, in the 3-qubit case, we consider the matrix element $\rho_{3,5}^{\mathrm{im}}$; its successive locations in $\pi_3$ are described by the sequence ``$2l3l4$'' which, according to Table~\ref{tab:rec}, corresponds to the local projector $\ket{RDV}$. The projector associated with $\rho_{4,9}^{\mathrm{re}}$ in a 4-qubit system, is instead determined by the locations in $\pi_4$ described by the sequence ``$2l3u12u$''; using Table~\ref{tab:rec}, one then finds $\ket{RRHD}$.

\begin{table}[!t]
  \begin{tabular}{c|c|c}
    Previous quadrant & New quadrant     & 1-qubit projector \\
    \hline\hline
    any       & 1 & $\ket{H}$  \\
    any       & 4 & $\ket{V}$  \\
    any except $2l$, $3l$ &  $2u$, $3u$  & $\ket{D}$  \\
    any except $2l$, $3l$ &  $2l$, $3l$  & $\ket{R}$  \\
    $2l$, $3l$     &  $2u$, $3u$  & $\ket{R}$  \\
    $2l$, $3l$     &  $2l$, $3l$  & $\ket{D}$ \\
    \hline
  \end{tabular}
    \caption{\label{tab:rec}
    Table determining the choice of the 1-qubit projector at each recursive step in Eq.~(\ref{eq:rec}). In the first step, ``previous quadrant'' is always ``any''. See text and Fig.~\ref{fig5:tQST-fig4-high-res} for details on the procedure.}
\end{table}

\subsection{\label{subsec:threshold}Choice of threshold}

The selection of an appropriate threshold value is dependent on the specific physical system used to implement the qubits ({\em e.g.}, noise level), the amount of available resources ({\em e.g.}, time requirements), and the desired quantum state to be generated.

In the case of IBMQ quantum processors considered here, it is possible to evaluate a convenient circuit-specific threshold using 
the IBMQ simulator available in the {\tt qiskit} package.
First, one simulates the unitary evolution of a ground-state initialized quantum register according to the circuit itself (we have used $n_s=10,000$ shots). Measuring all qubits yields the expected diagonal counts in the absence of errors, which can be separated into zero and non-zero counts. Second, one uses the IBMQ simulator (which includes the effect of noise) to run the circuit a number of times (100 in our case) and record: the maximum value of the counts among the expected zero elements of the diagonal, $c_0^{\mathrm{max}}$; and the minimum value of the counts for the smallest expected non-zero diagonal element, $c_{>0}^{\mathrm{min}}$. Third, one defines, in a conservative way: the {\it noise threshold} as $t_0 = c_0^{\mathrm{max}} + n\sqrt{c_0^{\mathrm{max}}}$; and the {\it signal threshold} as $t_{>0} = c_{>0}^{\mathrm{min}} - n\sqrt{c_{>0}^{\mathrm{min}}}$. The square root terms consider the variability of the counts $c_0^{\mathrm{max}}$, $c_{>0}^{\mathrm{min}}$ each time the circuit is simulated; the $n$ factor takes finally into account that, for the quantum processors considered, the noise increases with the number of qubits $n$. Then, we use as the circuit-specific (normalized) threshold the quantity
\begin{align}
t = \mathrm{max}(t_0, t_{>0})/n_s,
\end{align}
which discards those diagonal entries most affected by noise. For the depth 3 quantum circuits analyzed herein, we generally have $t_{>0}\gtrsim t_0$;
tQST works best whenever \mbox{$t_{>0}\gg t_0$}, whereas $t_{>0}\sim t_0$ indicates an unfavorable signal-to-noise ratio.

\subsection{\label{sec:fbound}Fidelity lower bound}

Fixing a threshold $t$ given a set of diagonal elements $\{\rho_{ii}\}$, establishes a lower bound on the fidelity achievable through a tQST reconstruction of $\rho$. Let ${\cal C}_t=\{\{ij\}\mid \sqrt{\rho_{ii}\rho_{jj}} < t\}$, and consider the estimator matrix $\rho_t=\{0 \quad \forall\ \{ij\}\in{\cal C}_t; \sqrt{\rho_{ii}\rho_{jj}} \ {\mathrm{otherwise}}\}$ (note that $\rho = \rho_t$ if $t=0$). Then, one has the inequalities~\cite{fuchs_vandegraaf,haah2016sample}
\begin{subequations}
    \begin{align}
		&1-\sqrt{F(\rho,\rho_t)}\leq \frac12\|\delta_t\|_1,\\
		&\|\delta_t\|_1\leq2\sqrt{{\mathrm{min}}(r(\rho),r(\rho_t))}\|\delta_t\|_2,
    \end{align}
\end{subequations}
with the fidelity defined according to $F(\rho,\rho_t)={\mathrm{Tr}} \left[ \sqrt{\sqrt{\rho}\rho_t\sqrt\rho} \right]$, $\|{\circ}\|_1$ the trace distance, $\delta_t=\rho-\rho_t$, and $r \left( \rho \right)$ the rank of the density matrix. Now, as $t$ increases, the state purity of $\rho_t$ decreases, so that $r(\rho)\le r(\rho_t)$. Indeed, the estimator $\rho_t$ is a density matrix with the off-diagonal elements below threshold set to zero. Physically this corresponds to making the density matrix more mixed and increasing its rank with respect to $\rho$. Additionally, $\|\delta_t\|_2\le\sqrt{\sum_{{\mathcal C}_t} \rho_{ii}\rho_{jj}}$. Thus, we obtain the lower bound
\begin{align}
		F_{\mathrm{bound}}(\{\rho_{ii}\}, t) =
        \left( 1-\sqrt{r(\rho)\sum_{{\cal C}_t}\rho_{ii}\rho_{jj}}
        \right)^2.
		\label{Flb}
\end{align}
 In our investigations, we noticed that the actual tQST fidelity can be significantly greater  than the one established by~(\ref{Flb}).

\subsection{\label{subsec:ml}Maximum likelihood reconstruction}

After measuring the counts $N_K$ associated with the density matrix elements $\vert \rho_K\vert> t$ using the corresponding projectors, the density matrix is reconstructed by minimizing the function
\begin{align}
	L&=\sum_K \left( \frac{n_K - N_K }{2 \sqrt{n_K}} \right)^2;&
	n_K = \bra{P_K}\rho\ket{P_K},
\end{align}
which is the negative log-likelihood function assuming that the noise on the counts has a Gaussian probability distribution~\cite{james_2001}. In order to minimize $L$, it is necessary to parametrize the density matrix $\rho$, in such a way that it fulfills the physical conditions of being Hermitian and positive semi-definite. A general approach is to write $\rho = T^\dagger T$ where $T$ is a triangular matrix. In this case, the number of parameters to be determined grows as $4^n$. 
If one has reasons to believe that $\rho$ describes a high-purity state, then one can express it as $\rho = V^\dagger V$ with ${\mathrm{dim}}(V)=r(\rho_t)\times 2^n$. In this case, the parameters needed for the reconstruction will scale more favourably with the system dimension.\\

\noindent{\large\bf Acknowledgments}\\

\noindent We thank Prof. Daniel F.V. James of the University of Toronto for useful discussions. 

\medskip
\noindent{\bf Funding:} This work is supported by PNRR MUR project PE0000023-NQSTI.

\medskip

\noindent{\bf Author contributions:} 
%
%
%
M.L. conceived the original idea. All authors developed the protocol. D.B. and G.G. developed the code and performed the experiments on IBMQ. All authors have contributed to writing the manuscript. D.B. and G.G. contributed equally to the work.

\medskip



\noindent{\bf Competing interests:} The authors declare that they have no competing interests.


\medskip
\noindent{\bf Data and materials availability:} The code implementing tQST is available upon reasonable request. All data needed to evaluate the conclusions in the paper are present in the paper and/or the Supplementary Materials.

\bibliography{new_tQST}

\end{document}


\title{Supplementary Material for "A Tailor-made Quantum State Tomography Approach"}


\author{Daniele Binosi}
\email{binosi@ectstar.eu}
\affiliation{European Centre for Theoretical Studies in Nuclear Physics
and Related Areas (ECT*, Fondazione Bruno Kessler); Villa Tambosi, Strada delle Tabarelle 286, I-38123 Villazzano (TN), Italy}

\author{Giovanni Garberoglio}
\email{garberoglio@ectstar.eu}
\affiliation{European Centre for Theoretical Studies in Nuclear Physics
and Related Areas (ECT*, Fondazione Bruno Kessler); Villa Tambosi, Strada delle Tabarelle 286, I-38123 Villazzano (TN), Italy}

\author{Diego Maragnano}
\email{diego.maragnano01@universitadipavia.it}
\affiliation{Dipartimento di Fisica, Universit\`a  di Pavia, via Bassi 6, 27100 Pavia, Italy}

\author{Maurizio Dapor}
\email{dapor@ectstar.eu}
\affiliation{European Centre for Theoretical Studies in Nuclear Physics
and Related Areas (ECT*, Fondazione Bruno Kessler); Villa Tambosi, Strada delle Tabarelle 286, I-38123 Villazzano (TN), Italy}
\author{Marco Liscidini}
\email{marco.liscidini@unipv.it}
\affiliation{Dipartimento di Fisica, Universit\`a  di Pavia, via Bassi 6, 27100 Pavia, Italy}

\maketitle

\section{\label{sec:ibmq_circuits}IBMQ circuits and tomography state experiments}

Random circuits were generated through the corresponding \verb|qiskit| command \verb|random_circuit| fixing the circuit depth to 3. We performed 10 independent runs for each circuit, conducing for each run a \verb|qiskit| state tomography  experiment, wherein the density matrix was reconstructed through a linear inversion of an over-complete set of $6^n$ measurements. This process generally yields a non-positive reconstructed state, which is subsequently rescaled to be positive semi-definite; the rescaling is achieved by computing the eigen-decomposition of the state and adjusting its eigenvalues using the method outlined in Ref.~\cite{Smolin:2011tob}.

Next for each circuit we fixed the threshold $t$ as discussed in Sec.~IVB, and proceeded to run the tQST reconstruction. Due to the presence of noise, different runs may result in different number of measurements to be performed, due to some of the diagonal elements becoming close to the threshold $t$. All results averaged over the 10 runs are shown in Tab.~\ref{tab:tQSTres}; results reported in Sec.~II correspond to the run with the minimum number of measurements performed. 

\begin{widetext}
\begin{center}
\begin{table}[!b]
    \begin{tabular}{
    >{\centering\arraybackslash}m{12mm}| 
    >{\centering\arraybackslash}m{10mm}|| 
    >{\raggedleft\arraybackslash}m{15mm}| 
    >{\raggedleft\arraybackslash}m{10mm}| 
    >{\raggedleft\arraybackslash}m{9mm}|| 
    >{\raggedleft\arraybackslash}m{11mm}| 
    >{\raggedleft\arraybackslash}m{15mm}| 
    >{\raggedleft\arraybackslash}m{15mm}| 
    >{\raggedleft\arraybackslash}m{15mm}| 
    >{\raggedleft\arraybackslash}m{10mm}| 
    >{\raggedleft\arraybackslash}m{10mm}| 
    >{\raggedleft\arraybackslash}m{10mm}| 
    >{\raggedleft\arraybackslash}m{10mm}  
    }
    \multicolumn{2}{c||}{} & \multicolumn{3}{c||}{\bf{Full QST}} & \multicolumn{7}{c}{\bf{tQST}} \\
    \cline{3-12}
    \hline
    Size & Circuit & \centering $F^{\mathrm{QST}}_{\mathrm{target}}$ &\centering $m$ &\centering $s$ &\centering $t$ &\centering $F_{\mathrm{bound}}$ &\centering $F^{\mathrm{tQST}}_{\mathrm{target}}$ &\centering $F^{\mathrm{tQST}}_{\mathrm{QST}}$ &\centering min $m$ &\centering max $m$ &\centering min $s$ & max $s$ \\
    \hline
    \hline
    \multirow{3}{*}{4-qubit} 
    & a) & 93.4(1)\% & 1,296 & 81 & 0.2077 & 47(1)\%& 93.5(4)\%& 96.9(3)\%& 28& 28& 9& 9\\
    & b) & 89.4(3)\% & 1,296 & 81 & 0.0348 & 56(3)\%& 89(1)\%& 95.2(9)\%& 72& 134& 27& 65\\
    & c) & 96.0(1)\% & 1,296 & 81 & 0.0074 & 93.4(5)\%& 95.4(3)\%& 98.0(2)\%& 190& 210& 79& 81\\
    \hline
    \hline
    \multirow{3}{*}{5-qubit} 
    & d) & 93.5(1)\% & 7,776& 243& 0.0182 & 70.8(6)\%& 93.4(3)\%& 94.8(4)\%& 130& 150& 69& 87\\
    & e) & 94.1(1)\%& 7,776& 243& 0.0269 & 57.4(7)\%& 94.0(3)\%& 95.9(3)\%& 272& 272& 81& 81\\
    & f) & 92.7(1)\%& 7,776& 243& 0.0066 & 85.7(3)\%& 92.5(6)\%& 96.3(3)\%& 456& 584& 187& 239\\
    \hline
    \hline
    \multirow{3}{*}{6-qubit} 
    & g) & 87.3(2)\%& 46,656& 729& 0.0127 & 59(2)\%& 86.9(5)\%& 90.0(6)\%& 556& 636& 229& 277\\& 
    h) & 85.1(1)\%& 46,656& 729& 0.0052 & 76(1)\%& 84.4(3)\%& 92.9(2)\%& 1404& 2322& 433& 637\\
    & i) & 86.8(1)\%& 46,656& 729& 0.0081 & 60.6(6)\%& 85.1(4)\%& 91.9(2)\%& 1804& 1986& 563& 585\\
    \hline
    \hline
    \multirow{3}{*}{7-qubit} 
    & l) & 89.0(1)\%& 279,936& 2,187& 0.0089 & 57(2)\%& 88.5(4)\%& 90.2(6)\%& 894& 1084& 295& 343\\
    & m) & 80.2(1)\%& 279,936& 2,187& 0.0050 & 64(1)\%& 79.2(3)\%& 87.7(2)\%& 4696& 5160& 1003& 1221\\
    & n) & 92.6(1)\%& 279,936& 2,187& 0,0053 & 61.2(6)\%& 92.2(2)\%& 92.6(2)\%& 2450& 2608& 963& 1067\\
    \end{tabular}
    \caption{\label{tab:tQSTres} Density matrix reconstruction of depth-3 random circuits on an IBMQ system. We grouped circuits according to their diagonal fillings, {\it i.e.}, the number of non-zero elements of the corresponding density matrix diagonal, for each qubit number: 25\% [a), d), g), l)] 50\% [b), e), h), m)], and 75\% [c), f), i), n)] for 4-, 5-, 6- and 7-qubit respectively. For each circuit, which was simulated 10 times, we report: for full QST, the (averaged) fidelity with respect to the target state, the number of measures, and the number of settings; for tQST, the circuit's threshold, the fidelity bound, the (averaged) fidelity with respect to the target state and the full QST reconstruction, the minimum and maximum number of measurements, and the minimum and maximum numbers of settings used to perform those measurements. All quoted uncertainties correspond to one standard deviation from the central value. In the 7-qubit case notice that it costs more to reconstruct circuit m) (which has 64 non-zero diagonal elements) than circuit n) (which has 96). This is due to a sizeable number of those elements falling below the threshold in the latter case, and therefore being excluded from the final measurement set.}
\end{table}

\end{center}	
\end{widetext}

\section{A comparison with compressed sensing quantum state tomography}

A pertinent question to consider is how the significant simplification in measurement complexity achieved by tQST compares with that obtained in compressed sensing (CS) QST, where the reconstruction of a rank-$r$ density matrix can be accomplished with ${\cal O}(r2^n\log^22^n)$ randomly chosen measurements~\cite{gross_2010}.

An experimental implementation of CSQST for a 7-qubit system, specifically a state of a `topological color code'~\cite{Bombin:2006sj} prepared in a trapped ion architecture, was reported in~\cite{riofrio2017experimental}; this system, described in some detail below, will be used for our comparison.

The 7-qubit states we want to analyze are given by
\begin{widetext}
\begin{align}
\ket{\overline{0}}&=\frac1{\sqrt8}\left( \ket{1010101}+\ket{1100011}+\ket{0101101}+\ket{0011011}+\ket{1001110}+\ket{0110110}+\ket{1111000}+\ket{0000000}\right),
\label{barzero}
\end{align}	
and
\begin{align}
\ket{\overline{1}}&=\frac1{\sqrt8}\left(\ket{0101010}+\ket{1010010}+\ket{0011100}+\ket{1100100}+\ket{0110001}+\ket{1001001}+\ket{0000111}+\ket{1111111}\right).
\label{barone}
\end{align}	
\end{widetext}
Denoting with $X_\ell$, $Y_\ell$ and $Z_\ell$ the Pauli operators acting on the qubit $\ell$, Eqs.~(\ref{barzero}) and (\ref{barone}) correspond to the eigenstates of the local operator $Z=Z_1Z_2Z_3Z_4Z_5Z_6Z_7$, with $Z\ket{\overline{0}}=\ket{\overline{0}}$ and $Z\ket{\overline{1}}=-\ket{\overline{1}}$; at the same time, they represent the positive eigenstates of all the six stabilizer operators $X_1X_2X_3X_4$, $X_2X_3X_5X_6$, $X_3X_4X_6X_7$, $Z_1Z_2Z_3Z_4$, $Z_2Z_3Z_5Z_6$, and $Z_3Z_4Z_6Z_7$. As a result, such a system represents the smallest possible instance of a topological color code, which encodes a single logical qubit $\ket{\overline{\psi}}$ (resilient to one bit flip, phase flip or a combined error of both) as an entangled state distributed over 7 physical qubits~\cite{Nigg:2014ded}.

Following~\cite{riofrio2017experimental}, for implementing CSQST we start by introducing a set of $m_S$ `Pauli basis measurements' (PBM) settings 
\begin{align}
	S^{(j)}&=V_1V_2V_3V_4V_5V_6V_7;& j&=1,\dots,m_S, 
\end{align}   
where the $V_\ell\in\{X_\ell,Y_\ell,Z_\ell\}$ are chosen at random, and the $\ell^{\mathrm{th}}$ qubit is measured in the basis of $V_\ell$. Each setting $S^{(j)}$ has then $2^7$ possible outcomes corresponding to the number of ways of picking one of the two eigenvectors of each of the Pauli matrices appearing in it. The original `Pauli correlators measurements' of~\cite{gross_2010} are then recovered from the relation
\begin{align}
	\mathrm{tr}\,(\rho S^{(j)})&=\sum_{k=1}^{2^7}(-1)^{\chi(k)}\bra{v^{(j)}_k}\rho\ket{v^{(j)}_k},
	\label{mel}
\end{align}
where $\ket{v_k^{(j)}}$ is the tensor product of the eigenvectors of $S^{(j)}$ associated to a specific outcome $k$, and $\chi(k)$ denotes the parity of the binary representation of the integer $k$. Clearly, PBMs provide very detailed information per setting.

In~\cite{riofrio2017experimental} the number of settings $m$ was chosen to be $127$, and for each setting the measurement was repeated $100$ times. As the number of repetitions is smaller than the number of possible outcomes, the measurements are undersampled resulting in a quite poor signal to noise ratio. In any case, according to Eq.~(\ref{mel}) one ends up measuring~$2^7$ matrix elements for each setting, or~\mbox{$m_S\times2^7=16,256$} matrix elements in total. 

For the tQST reconstruction of the states $\ket{\overline{0}}$ and $\ket{\overline{1}}$ mentioned earlier, only $2^7 + 8 \cdot 7 = 184$ matrix elements need to be measured. This is nearly two orders of magnitude less than the requirements for CSQST. However, in IBMQ or ion-trap architectures, Pauli basis measurements naturally emerge: repeatedly measuring all qubits following a setting $S^{(j)}$, one gains access to the statistical distribution of all $2^7$ possible outcomes associated with that setting. Thus, in these platforms the cost of obtaining the entire set of measurements (CSQST) or a single one (tQST) is, for all practical purposes, equivalent, and the efficiency of the QST method employed is rather gauged by the number of settings $m_S$.


For such platforms, the tQST framework provides an upper bound on the number of settings needed for state tomography: these are given by the settings that contain the states which provide information on the matrix elements $(i,j)$ that have to be measured, since from each projectors determined as in Sec.~IV A, one can immediately deduce the setting that contains it. Hence, the number of settings is at most $1+N(N-1)$, where $N$ is the number of above-threshold elements measured along the diagonal in the computational basis (which corresponds to the PBM setting where $V_\ell = Z_\ell,\ \forall\ \ell=1,\dots,7$). In practice, many of the PBM settings found in this way turn out to be the same, resulting in a further reduction. In the case of the color-code states $\ket{\overline{0}}$ or $\ket{\overline{1}}$, this procedure picks out 57 settings, which are less that the $127$ (random) settings used in \cite{riofrio2017experimental}. We hasten to emphasize that, at variance with CSQST where the settings are sampled randomly, the tQST procedure {\em determines} the optimal settings, and guarantees that they provide enough information to reconstruct the quantum state.

\begin{center}
\begin{table}[!t]
    \begin{tabular}{>{\centering\arraybackslash}m{13mm}|>{\raggedleft\arraybackslash}m{23mm}|>{\raggedleft\arraybackslash}m{20mm}}
    \multicolumn{1}{>{\centering\arraybackslash}m{13mm}|}{State} 
    & \multicolumn{1}{>{\centering\arraybackslash}m{23mm}|}{N. of settings} 
    & \multicolumn{1}{>{\centering\arraybackslash}m{20mm}}{Fidelity}\\
    \hline
    \hline
    \multirow{3}{*}{$\ket{\overline{0}}$} & 127 (random) & $98(1)\%$\\
    & 57 (random) & $92(3)\%$ \\
    & 57 (tQST) & $99.4(1)\%$ \\
    \hline
    \multirow{3}{*}{$\ket{\overline{1}}$} & 127 (random) & $98(1)\%$\\
    & 57 (random) & $92(3)\%$ \\
    & 57 (tQST) & $99.4(1)\%$ \\
    \hline
    \end{tabular}
    \caption{\label{tab:CS}Comparison of compressed sensing reconstruction of the anticipated states $\ket{\overline{0}}$ and $\ket{\overline{1}}$ when using: $127$ PBM settings randomly chosen, as originally done in~\cite{riofrio2017experimental}; $57$ PBM settings, again randomly chosen; and the $57$ PBM settings obtained through tQST. In all cases we use 100 shots, with fidelity's averages and standard deviations evaluated over 10 reconstructions. Evidently, tQST PBM settings provide a better choice over a random selection.}
\end{table}
\end{center}

In Table~\ref{tab:CS} we report the results of the reconstruction of the $\ket{\overline{0}}$ and $\ket{\overline{1}}$ `anticipated states' ({\it i.e.}, the states that an ideal experiment would prepare) using the 57 tQST-determined PBM settings and 100 shots as in \cite{riofrio2017experimental}. In this case we obtain an almost perfect reconstruction, with fidelity in excess of 99\%. The quality of these settings can be appreciated by the fact that using 57 random settings,  results in significantly smaller fidelities. Even using 127 random settings (as per the original CSQST approach of~\cite{riofrio2017experimental}) one does not quite reach the extremely good reconstruction of the 57 tQST settings.

These results strongly suggest that considerations based on tQST  could be used to optimize the settings required for CSQST. The idea is to replace random sampling of settings with a deterministic procedure based on an initial measurement along the computational basis. We are actively exploring this research direction and anticipate providing a comprehensive report on our findings~soon.

\bibliography{suppl_new_tQST}